\documentstyle[mncite]{mn}

\newif\ifAMStwofonts
\def\fa{{151 {\rm \thinspace MHz}}}
\def\fb{{178 {\rm \thinspace MHz}}}
\def\fc{{1.4 {\rm \thinspace GHz}}}
\def\fd{{2.7 {\rm \thinspace GHz}}}
\def\fe{{5 {\rm \thinspace GHz}}}

\def\fg{{750 {\rm \thinspace MHz}}}

\ifoldfss
  \ifCUPmtlplainloaded \else
    \NewTextAlphabet{textbfit} {cmbxti10} {}
    \NewTextAlphabet{textbfss} {cmssbx10} {}
    \NewMathAlphabet{mathbfit} {cmbxti10} {} % for math mode
    \NewMathAlphabet{mathbfss} {cmssbx10} {} %  "   "    "
  \fi

  \ifAMStwofonts
    \ifCUPmtlplainloaded \else
      \NewSymbolFont{upmath} {eurm10}
      \NewSymbolFont{AMSa} {msam10}
      \NewMathSymbol{\upi}     {0}{upmath}{19}
      \NewMathSymbol{\umu}     {0}{upmath}{16}
      \NewMathSymbol{\upartial}{0}{upmath}{40}
      \NewMathSymbol{\leqslant}{3}{AMSa}{36}
      \NewMathSymbol{\geqslant}{3}{AMSa}{3E}

      \let\leq=\leqslant \let\le=\leqslant
      \let\geq=\geqslant \let\ge=\geqslant
    \fi
  \fi

\fi % End of OFSS
\ifnfssone
  \newmathalphabet{\mathit}
  \addtoversion{normal}{\mathit}{cmr}{m}{it}
  \addtoversion{bold}{\mathit}{cmr}{bx}{it}
  \newmathalphabet{\mathbfit} % math mode version of \textbfit{..}
  \addtoversion{normal}{\mathbfit}{cmr}{bx}{it}
  \addtoversion{bold}{\mathbfit}{cmr}{bx}{it}
  \newmathalphabet{\mathbfss} % math mode version of \textbfss{..}
  \addtoversion{normal}{\mathbfss}{cmss}{bx}{n}
  \addtoversion{bold}{\mathbfss}{cmss}{bx}{n}
  \ifAMStwofonts
    \ifCUPmtlplainloaded \else
      \UseAMStwoboldmath
      \makeatletter
      \new@mathgroup\upmath@group
      \define@mathgroup\mv@normal\upmath@group{eur}{m}{n}
      \define@mathgroup\mv@bold\upmath@group{eur}{b}{n}
      \edef\UPM{\hexnumber\upmath@group}
      \new@mathgroup\amsa@group
      \define@mathgroup\mv@normal\amsa@group{msa}{m}{n}
      \define@mathgroup\mv@bold\amsa@group{msa}{m}{n}
      \edef\AMSa{\hexnumber\amsa@group}
      \makeatother
      \mathchardef\upi="0\UPM19
      \mathchardef\umu="0\UPM16
      \mathchardef\upartial="0\UPM40
      \mathchardef\leqslant="3\AMSa36
      \mathchardef\geqslant="3\AMSa3E

      \let\leq=\leqslant \let\le=\leqslant
      \let\geq=\geqslant \let\ge=\geqslant
    \fi
  \fi
\fi % End of NFSS release 1
\ifnfsstwo
  \DeclareMathAlphabet{\mathbfit}{OT1}{cmr}{bx}{it}
  \SetMathAlphabet\mathbfit{bold}{OT1}{cmr}{bx}{it}
  \DeclareMathAlphabet{\mathbfss}{OT1}{cmss}{bx}{n}
  \SetMathAlphabet\mathbfss{bold}{OT1}{cmss}{bx}{n}
  \ifAMStwofonts
    \ifCUPmtlplainloaded \else
      \DeclareSymbolFont{UPM}{U}{eur}{m}{n}
      \SetSymbolFont{UPM}{bold}{U}{eur}{b}{n}
      \DeclareSymbolFont{AMSa}{U}{msa}{m}{n}
      \DeclareMathSymbol{\upi}{0}{UPM}{"19}
      \DeclareMathSymbol{\umu}{0}{UPM}{"16}
      \DeclareMathSymbol{\upartial}{0}{UPM}{"40}
      \DeclareMathSymbol{\leqslant}{3}{AMSa}{"36}
      \DeclareMathSymbol{\geqslant}{3}{AMSa}{"3E}

      \let\leq=\leqslant \let\le=\leqslant
      \let\geq=\geqslant \let\ge=\geqslant
    \fi
  \fi
\fi % End of NFSS release 2

\ifCUPmtlplainloaded \else
  \ifAMStwofonts \else % If no AMS fonts
    \def\upi{\pi}
    \def\umu{\mu}
    \def\upartial{\partial}
  \fi
\fi

\title[Radio source unification and evolution]
{Extragalactic radio source evolution under the dual-population
unification scheme}

\author[C.A. Jackson \& J.V. Wall]
       {C.A.Jackson$^1$\thanks{Current address Department of
Astrophysics, School of Physics,
       University of Sydney, NSW 2006, Australia} \& J.V.Wall$^2$ \\
       $^1$Institute of Astronomy, University of Cambridge, Madingley Road,
               Cambridge, CB3 OHA, UK\\
       $^2$Royal Greenwich Observatory, Madingley Road,
               Cambridge, CB3 OEZ, UK}

\date{Accepted 1998 ??? ??.
      Received 1998 ??? ??;
      in original form 6 Aug 1998}

\pagerange{\pageref{firstpage}--\pageref{lastpage}}
\pubyear{1998}

\begin{document}
\maketitle
\label{firstpage}
\begin{abstract}

We show that a dual-population unification scheme provides a successful
paradigm with which to describe the evolution and beaming of all bright
extragalactic radio sources. The paradigm consists of two intrinsic
radio-source populations, based on the two distinct radio-galaxy
morphologies of Fanaroff-Riley classes I and II. These represent the
`unbeamed' or `side-on' parent populations of steep radio spectra; the
`beamed' source types including flat-spectrum quasars and BL\,Lac objects,
arise through the random alignment of their radio-axis
to our line-of-sight where Doppler-beaming of the relativistic
radio jets produces highly anisotropic radio emission.

We develop the model in two stages. In the first stage the source space
density as a function of cosmic epoch is determined for the two parent
populations, and for this we use low-frequency source-count and
identification data to avoid biases due to Doppler-enhanced radio
emission. The second stage defines the beaming models for each population,
using high-frequency survey data and in particular the 5-GHz source count
in which at high flux densities the flat- and steep-spectrum sources
contribute in similar measure.  We assume that the flat-spectrum objects,
quasars and BL\,Lac objects are `beamed' versions of FRI and FRII objects
in which the close alignment of radio-axis with line-of-sight has changed
the radio appearance into a core-dominated (flat-spectrum) object.  We
adopt a simple parameterisation of the beaming, orient the parent
populations at random with a Monte-Carlo process, and use a minimisaton
process to determine beaming parameters which yield a best fit to the
5-GHz souce count. The best-fit parameters are found to be in good
agreement with those measured observationally for individual radio
sources. In this, the model accurately reproduces the change in
source-count form with frequency. Indeed the unified-scheme paradigm has
great predictive power,  and we show how the model successfully describes
several additional and independent data sets. 

\end{abstract}


\section{Introduction}

The dramatic space density evolution of radio-loud AGNs has
been shown to mirror the cosmic star formation rate
\cite{wal98,boy98,sha98}.
Similar strong epoch dependence is seen in the faint 
blue galaxy population \cite{met95,roc96}, IRAS galaxies 
\cite{sau90} and 
the radio-bright starbursting galaxies \cite{con84,row93}. 
These correlated epoch-dependences provide enticing clues
towards understanding AGN physics, in the context of triggers
to activity such as coalescence of critical
mass, merging, fuel supply, and the lobe confinement provided by an
evolving IGM.

Our concept of the radio-source populations developed slowly.  The
work of
\scite{lon66} and collaborators \cite{dor70} demonstrated conclusively the
dependence of the radio-AGN evolution on radio luminosity: at redshifts
$>1$ the most powerful radio sources showed space-density enhancements by
factors of $>10^3$ over present-day densities, whereas the less-powerful
sources showed little or no enhancement.  Subsequently \scite{fan74}
demonstrated a morphological dichotomy: double-lobe radio sources with
regions of highest surface brightness close to the nucleus (Fanaroff-Riley
type (FRI)) have radio luminosities consistently lower than those with
highest surface brightness at the extremities (FRII).  The FRI -- FRII
division occurs approximately\footnote{Note that this division is {\it
not} a simple function of radio power and we return to this point
in section 3.4.1.} at a radio power of $\log_{10}(P_\fa) \sim$
10$^{25}$ W Hz$^{-1}$ sr$^{-1}$, close to the luminosity dividing
strongly-evolving radio sources from those showing little evolution. On
the basis of this coincidence, \scite{wal80b} suggested that the FRI
population showed near-constant space density while the FRII objects
evolved strongly with cosmic epoch. 

The early surveys were at relatively low frequencies and catalogued
predominantly steep-spectrum radio galaxies of extended
double-lobed structures.
During the late 1960s and the 1970s surveys at
frequencies $>2$ GHz found a large number of radio AGN whose spectra did
not decrease monotonically with frequency - sources of hard radio spectra,
described (unfortunately) as `flat spectrum' objects.  In contrast to the
host objects of steep radio spectra, predominantly giant ellipticals, the
majority of the flat-spectrum objects were identified with QSOs (quasars)
or BL\,Lac objects and had compact radio structures, unresolved until
mapped with VLBI. \scite{sch76} and \scite{mas77} claimed that despite
their apparent high radio luminosity of such objects, they showed little
evolution. This result was disputed and eventually the comprehensive 
analysis of
\scite{dun90} showed that qualitatively the flat-spectrum population
showed a similar evolutionary behaviour to the steep-spectrum population: 
luminosity-dependent density evolution was required. However, at a
particular radio luminosity, the space density evolution of the
flat-spectrum population is {\it less} than that of the steep-spectrum
population.  At the time, physical connection between the
steep-spectrum and flat-spectrum populations were not considered in such
analyses.

It was the discovery of relativistic expansion \cite{coh71,mof72} through
repeated VLBI measurements of the brighter `flat-spectrum' sources which
laid the foundation for unification paradigms. Whilst relativistic motion
is the primary mechanism giving rise to anistropic radio emission in
radio-loud AGN, a second mechanism, the presence of a dusty molecular
torus \cite{ant85} was invoked to explain the optical emission
characteristics of some sources.  From these two physical mechanisms the
paradigm emerged in which the random orientation of a `parent' population
on
the plane of the sky gave rise to radio sources of apparently diverse
type.  The {\it unified schemes} proposed for radio-loud AGN
\cite{sch79,orr82,sch87,bar89} have both radio and optical appearance
governed by proximity of the line-of-sight to the radio (presumably
rotation) axis. In the radio regime, close coincidence changes the
morphology from an extended steep-spectrum source to a source with a
dominant
flat-spectrum `core' from the Doppler-boosted base of the foreground jet.
The optical appearance changes from galaxy as viewed side-on to quasar or
BL\,Lac object when orientation permits a view into the torus opening
angle and optical/UV radiation from the accretion disk 
system dominates stellar emission.

If we are to understand the physical significance of the evolution which
originally suggested different populations, it is essential to delineate
the populations on a physical basis. We cannot map the evolution
meaningfully until we know the physically-distinct populations for which
we wish to derive it. 

The primary purpose of this paper is to advance this chicken-and-egg
situation in the following way. Firstly we consider the dual-population
unified scheme, based on the premise that the two parent populations are
FRI and FRII radio galaxies, which give rise to BL\,Lac objects and
quasars as beamed progeny.  Our basic assumption is that {\it all} radio
sources detected above $S_\fe \sim$ 3 mJy are encompassed by this scheme.
Using the best available data in terms of source counts and
identifications at low frequencies, we estimate the space densities of the
parent populations.  We then set up the unified models to define critical
angles within which the line-of-sight turns these lobe-dominated sources
into core-dominated sources. Using a Monte Carlo orienting process we
allow free rein to the fitting process to reproduce a high-frequency radio
source count.  This process results in a model for the Doppler beaming of
the radio jets with physical parameters which closely mirror those
observed in VLBI monitoring surveys. The dual-population unified scheme
makes numerous quantitative predictions which can be tested, and a number
are described in this paper. Crucial further observations are delineated.

This is not the first attempt to incorporate a unification paradigm within
space density analyses. Before the discovery of dusty tori, \scite{sch79}
proposed a model of superluminal expansion for quasars with a view to
explaining the ratio of radio-quiet QSOs to radio-loud quasars. A second
seminal study by \scite{orr82} proposed that the core-dominated quasars
were aligned versions of steep-spectrum quasars based on the twin
relativistic-jet model of \scite{bla74}.  Using the
observed
redshift distribution of 3C quasars and a simple beaming model they
calculated the Lorentz factor, $\gamma$, from the observed core-to-lobe
flux ratio distribution, $R$, for a sample of quasars.  They found that
the flat-spectrum quasar fractions at 2.7 GHz and 5 GHz could be
reproduced by single values of $R_{T}$ and $\gamma$.  These views of
unification were limited, primarily in that the dusty torus did not form
part of the modelling.  More recently, Urry and Padovani (1995 and
references therein)  considered a simple two-population unified scheme,
where FRII radio galaxies are the parent population of {\it all} quasars
and FRIs are the parent population of {\it all} BL Lac-type objects.  They
fitted the observed luminosity function for the steep-spectrum,
`misaligned' objects to the flat-spectrum, `aligned' luminosity function,
using a pure luminosity evolution model coupled with simple model of the
Doppler beaming which required a large range of Lorentz factors.

In contrast, this paper starts with a definition of the two parent
populations based on low-frequency data alone, 
and derives independent space-density descriptions for each. Using
high-frequency data it then
derives beaming models for each population. An embryonic
version of the present model was described by \scite{wal97}; the present
paper adopts the same thesis and procedure, presents the data in detail,
describes a refined version of evolution for the
parent
population and develops beaming models which provides a {\it
distribution} of apparent jet velocities as well as core-to-lobe flux
ratios. In addition, cosmological tests of the paradigm are presented. 

Throughout this paper we use $h$=0.5 and $\Omega_0$=1.0.

\section{The Dual-Population Unified Scheme}

\label{dualp}

The current unification paradigm for radio-loud AGN is based on two
`parent' populations, namely (i)  the high-radio-power FRII galaxies and
(ii) the moderate-radio-power FRI galaxies.  Both populations exhibit
anisotropic radiation arising from superluminal motion (Doppler beaming) 
of the radio jets. In addition, obscuration by a dusty torus contributes
to the orientation-dependent appearance of the high-power FRIIs; when the
torus/rotation axis and the line-of-sight are approximately aligned,
radiation from the accretion disk about the black hole 
dominates the galaxy emission.
The orientation angle of the radio axis on the plane of the sky determines
the observed characteristics and the hence the classification of the
object. 

In the case of the FRII radio galaxy, a side-on view reveals a
lobe-dominated steep-spectrum radio source hosted by a giant elliptical
galaxy, possibly with narrow emission lines of high excitation. As the
observer's line of sight comes closer to the radio axis, the object
appears as a `steep-spectrum quasar', the line-of-sight bringing the blue
light / broad-line region of the nuclear accretion disk into view to
dominate
the light of the host galaxy. At close coincidence, the radio radiation 
becomes dominated by the Doppler-enhanced
core or base of the forward jet; features may show superluminal 
velocities. In
the case of the FRI radio galaxies,
the side-on view shows a galaxy of weak-to-no narrow emission lines
while the on-axis view reveals a BL\,Lac object of
essentially featureless optical continuum and compact radio structure
showing superluminal velocities.

The steep-spectrum side-on objects, FRI and FRII radio galaxies plus
steep-spectrum quasars, dominate the results from low-frequency surveys,
the complete samples and source counts. High-frequency radio surveys
favour the
radio sources of harder spectrum, {\it i.e.} beamed, core-dominated sources,
BL\,Lac objects and core-dominated quasars. Raising survey frequency 
selects samples and source-counts dominated by objects with their radio
axes close to our line-of-sight. 

The hypothesis that lobe-dominated steep-spectrum and core-dominated
flat-spectrum quasars are increasingly aligned versions of FRII radio
galaxies is that originally proposed by \scite{sch87,pea87,bar89}.  The
corresponding scheme for FRI - BL\,Lac unification was put forward by
\scite{bro83,per84}.  However, there is accumulated evidence that the
unified scheme is not quite so straightforward. In particular the FRII
sources are {\it not} a homogeneous population in terms of their
optical/UV spectral characteristics:  \scite{hin79} showed that FRIIs can
have strong narrow optical/UV emission lines (class `A') or only weak, if
any, lines, similar to FRI sources (class `B').  \scite{lai94} showed that
in the 3CRR sample of \scite{lai83}, this dichotomy was reflected by a
clean division between high-excitation / strong-emission line objects and
low-excitation / weak-to-no-emission-line objects. This dichotomy in the
FRII population is reflected in the `dual population' unified scheme which
we propose.  We suggest that {\it only the class `A' FRII sources} appear
as lobe-dominated broad-line quasars, or core-dominated broad-line quasars
at small angles of their radio axes to our line of sight. 
Moreover, it is only for the class `A' sources that the presence
of an obscuring torus is invoked.

The role of the low-excitation FRIIs has not been considered in unified
models or space-density analyses to date.  However the shape of the radio
luminosity function for the FRII population shows that there must be many
more low-excitation sources than high-excitation ones, given that
excitation status correlates with radio power \cite{lai94,bar94}.  If
these objects have relativistically-beamed radio jets then they must show
a beamed sub-population. There is strong evidence, both direct and
circumstantial, that
this sub-population is seen as BL\,Lac objects. Firstly, around some
BL\,Lacs there are extended
structures consistent in scale and power with the lobes of FRII sources
\cite{bro82,kol92,mur93,dal97}. Secondly, there are too
few FRIs
for them to be the only class of host galaxy for BL\,Lacs \cite{owe96}. 
Thirdly the lack of observed narrow emission lines in the
parent
sources, in the class-B FRII galaxies, correlates with the optical
spectral
properties of BL\,Lacs; the majority of radio galaxies of intrinsic radio 
luminosities more powerful than the FRI/FRII divide, show passive
elliptical-galaxy spectra \cite{rix91}. 

Thus our dual-population unified scheme, illustrated in Figure 1, consists
of 
two parent populations: (a) the 
high-radio-power FRII radio galaxies which are the
parents of {\it all} radio quasars and {\it some} BL\,Lac-type
objects, and (b) the moderate-radio-power FRI radio 
galaxies which are the parents of BL\,Lac-type objects.

\begin{figure*}
\vspace{9.0in}
\includegraphics{diagus.figps}
\end{figure*}

Whilst the identified classes of extragalactic radio sources are
diverse (see {\it e.g.} \pcite{wal94} for a detailed discussion),
we adopt the working hypothesis that all radio source types found in
surveys between 0.15 and 10 GHz down to $S_\fc \sim$1 mJy 
belong to one of the
two parent populations or to one other population, the 
starburst galaxies \cite{con84,win85}. 
Table~\ref{ers} describes types of extragalactic radio source commonly
discussed
in the literature;
the only notable exclusion is cluster relic sources \cite{gio93} which are
so few in number and so low in space density as to have minimal impact on
our analysis.

\begin{table*}
\caption{Extragalactic radio source classes.}
\label{ers}
\begin{tabular}{lccccc}
 & UV/optical & Radio &  & \multicolumn{2}{c}{Contribution} \\
 & emission-line & spectrum & & \multicolumn{2}{c}{to source count} \\
Population & signature
 & $\nu\sim$5 GHz & Class & $\nu\stackrel{<}{\sim}$1 GHz 
& $\nu\stackrel{>}{\sim}$1 GHz \\
\hline
\\
FRII, class A$^{\flat}$ & narrow & steep & RG & prominent & prominent \\
                         & broad & steep & RG$^{\sharp}$ & prominent & prominent \\
                         & narrow & `flat' &  quasar & $<$5\% & prominent \\
\\

FRII, class B$^{\flat}$  & weak/none & steep & RG & prominent & prominent \\
                      & none & `flat' & BL Lac & $<$5\% & $<$10\% \\
\\

FRI & weak/none & steep & RG & prominent & prominent \\
       & none & `flat' & BL Lac & $<$5\% & $<$10\% \\
\\
 
Starburst & narrow & steep & galaxy & $<$1\% & $<$1\% \\
\\

GPS  & narrow & `flat' & RG & $<$1\% & $\sim$5\% \\
     & broad & `flat' & quasar & $<$1\% & $\sim$5\% \\
\\

CSS  & narrow & steep & RG & $\sim$15\% & $<$1\% \\
     & broad & steep & quasar & $\sim$5\% & $<$1\% \\
\\
\hline
\end{tabular}
\\
\vspace*{0.1in} 
 
{\it $\flat$ Hine \& Longair (1979) emission-line class.} \\
{\it $\sharp$ alternatively a steep-spectrum quasar.}
\end{table*}

{\bf The FRI and FRII sources} are the powerful radio sources
explicity described by the dual-population unified scheme of
 the previous section.  
\\
\indent {\bf The starburst galaxies}
are late-type, in contrast to the other populations
 in the table.  They have intrinsically-weak radio emission, and 
dominate radio samples at low flux densities \cite{win85}. 
There is no evidence of any Doppler beaming of their radio emission;
they lie outside the dual-population unified scheme. 
The contribution of this population
is included in our analysis using previously-determined luminosity
functions and evolution. \\
\indent {\bf The GPS and CSS} are `peaked-spectrum' sources,
objects with peaks in their radio spectra between 0.1 and 3 GHz.
There is accumulating
evidence that  GigaHertz-Peaked Spectrum and Compact Steep-Spectrum
(GPS and CSS) sources
represent young stages ($<10^4$ yr) of FRII radio sources 
\cite{rea95,fan96}. 
Therefore these source types are encompassed in the dual-population unified
scheme given that (i) the frequency of
broad-line obects in these classes is in general agreement with the
simple model of a broad-line region being obscured by a surrounding torus
\cite{fan90} and (ii) superluminal motion
has been observed in nearby GPS galaxies, with Lorentz factors
similar to those inferred for classical FRIIs
\cite{gio95,tay95}.  Moreover the shortfall
of peaked-spectrum quasars may due to
Doppler-boosting of the forward radio jet flattening
the characteristic peaked radio spectra with the result that
GPS and CSS quasars are often categorized as `normal' FRII quasars
\cite{sne97}.

\section{Radio source space densities within the unified scheme}

\subsection{Parametric models of radio source evolution}

The first stage in our analysis is to determine the space density
evolution of the two parent populations, allowing each to undergo
quite separate evolution histories. For each of the two
populations we determine its local radio luminosity function
and adopt a simple parametric description of its evolution
with epoch. Whilst the parametric approach has the disadvantage of applying a
rigid form of evolutionary behaviour, it has the advantage of 
comprising only a limited set of parameters ({\it e.g.} 
\pcite{wal80a,orr82,mor87,row93}).
The alternative approach is a free-form fit 
\cite{rob80,pea81,dun90},
advantageous as there is no 
requirement to `guess' the form of the
evolution prior to determining an acceptable fit. However the free-form
method yields a large parameter set which is difficult to manage in 
subsequent analyses and which does not necessarily lead to further
physical understanding. 
A final consideration in our choice of approach is that in describing
independent evolution histories for the FRI and FRII
populations, we are limited to datasets which are simply too 
small to use in a free-form analysis.

There are two distinct types of parameterized evolution which may be
applicable
to extragalactic radio sources. In the case of
{\it luminosity evolution} (LE) the radio luminosity of
sources change with epoch.  If all sources undergo the
same degree of LE it is described `pure luminosity evolution' (PLE).
Alternatively in {\it density evolution} (DE) 
the co-moving space density of a population
changes with epoch. If all sources undergo the same degree of density
evolution it is described as `pure density evolution' (PDE).  A 
combination of luminosity and density evolution is of course possible.
Previous analyses which attributed a single evolution
history to {\it all} radio sources ({\it i.e.} combining the FRI and 
FRII populations into a single `steep-spectrum' population) 
have concluded that neither PLE nor 
PDE are applicable ({\it e.g.} \pcite{wal80a}) 
and instead successful descriptions of the data have always required
a form of luminosity-dependent evolution.

Furthermore there are two popular forms of parametric evolution,
applicable to
both density and luminosity evolution, which may 
have some physical significance: {\it (i)}
power-law evolution of the form $(1 + z)^{\eta}$, $(1+z)^{-1}$ being the
universal scale factor, and {\it (ii)} exponential evolution of
the form
$\exp M \tau(z)$, $\tau$ being the look-back time.  Whilst power-law
evolution was successfully applied to models for optically-selected
QSOs \cite{boy88}, exponential 
evolution has been found to be more applicable
for radio-selected samples, as it
reproduces the strong cosmic evolution observed at relatively modest
redshift ({\it e.g.} \pcite{dor70}). 

\subsection{Methodology}

\label{emethod}

Our approach follows that of \scite{wal80a} by
which the local radio luminosity function is determined
given a complete {\it luminosity distribution} and a chosen {\it evolution
function} $F$. The evolution function $F(P,z)$
modifies the local radio luminosity function, $\rho_0$, 
to give the radio luminosity
function at any epoch $z$, {\it i.e.} $\rho(P,z) = \rho_{0}(P)
F(P,z)$. 
This description is entirely general and can describe luminosity
evolution, density evolution, or a mixture. 

The steps in this process to a description of space density are:\\

\hspace*{0.08in}{\it For each population in the source count:} 

\begin{enumerate}

\item  Compile the luminosity distribution $N(P)$
from a complete flux-limited sample of identifications and redshifts.

\item Choose an evolution function $F(P,z)$. 

\item Calculate the local radio luminosity function $\rho_{0}(P)$: \\

\begin{equation}
\rho_{0}(P) dP = \frac{N(P) dP}{\int_{0}^{z(S_{0})} F(P,z) dV(z)} \\
\label{eq:rhono}
\end{equation}
\noindent
so that the radio luminosity function  $\rho(P,z) = \rho_{0}(P) F(P,z)$ 
is then known at all redshifts.

\item Calculate the contribution to the radio source count: 

\begin{equation}
N(\ge S) = \int_{0}^{\infty} dP \int_{0}^{z(S)} \rho_{0}(P) F(P,z) dV(z) \\
\label{eq:count}
\end{equation}
\noindent
where $z(S)$ is the redshift at which a source of intrinsic radio power $P$
contributes at flux density $S$.   A redshift cutoff $z_{c}$ may be
imposed, beyond which $F = 0$. In practice the differential source count
for the population
is calculated, flux-binned for comparison purposes exactly as
the observed count has been compiled. \\

\hspace{0.08in}{\it When the contributions from all populations  have
  been found:} \\

\item Sum the single-population source counts to give the total predicted
count.
The predicted and observed differential source counts are compared and the
evolution functions are adjusted, the process iterating
from step (ii) until an acceptable fit to the observed count
is obtained.

\end{enumerate}

\subsection{\boldmath Determining the evolution functions $F(P,z)$}

A suitable trial form for the evolution function can be determined from
the  $<\!\!V/V_{max}\!\!>$ statistic for a complete sample.  The 
best sample available at low frequencies is the 3CRR sample at
178~MHz (\pcite{lai83} and more
recently published data collated by R. Laing, private communication)
as it is complete in terms of radio morphologies and redshifts.

\addtocounter{figure}{1}

\begin{figure}
\vspace{3.0in}
\includegraphics{3crvvm.figps}
% from cam_home/cam_save_prgs\save_prgs/vvmtest2
\caption{$V/V_{max}$ values ($\ast$) for the 
137 steep-spectrum FRII sources in 3CRR. 
$<V/V_{max}>$ for the total sample is shown dashed.
$<V/V_{max}>$ values for bins in $\Delta \log_{10}(P_\fa)$=0.5 are
shown as data points ({\Large +}).}
\label{3crvvm}
\end{figure}

\begin{enumerate}

\item {\bf FRII Evolution.} A plot of $V/V_{max}$ (Figure~\ref{3crvvm})
for the 137 steep-spectrum FRII sources in 3CRR
indicates that the highest radio power FRII 
sources have undergone far more
evolution than the lower-power ones.
Whilst $<V/V_{max}>$ = 0.664 ($\sigma$=0.025) for the
137 FRII sources, this value rises from 0.415 ($\sigma$=0.118)
at $\log_{10}(P_\fa)$ = 24.25
to 0.807 ($\sigma$=0.052)
at $\log_{10}(P_\fa)$ = 28.25. 

\item{\bf FRI Evolution.}
The $V/V_{max}$ statistic for the 26 FRI sources in 3CRR is
shown in Figure~\ref{3crvvm1}. In contrast to the strong evolution
indicated for the FRII population, the FRIs show little
evidence of evolution, with
$<V/V_{max}>$ = 0.314 ($\sigma$=0.057) for all 26 sources
possibly reflecting some negative evolution.  The
highest-power FRI sources in 3CRR have a value of
$<V/V_{max}>$ = 0.507 ($\sigma$=0.144)
at $\log_{10}(P_\fa)$ = 25.75.

\end{enumerate}

\begin{figure}
\vspace{3.0in}
\includegraphics{3crvvm1.figps}
% from cam_home/cam_save_prgs/vvmtest1
\caption{$V/V_{max}$
values ($\ast$) for the 26 steep-spectrum FRI sources in 3CRR. 
$<V/V_{max}>$ for the total sample is shown dashed. 
$<V/V_{max}>$ values for bins in $\Delta \log_{10}(P_\fa)$=0.5
are
shown as data points ({\Large +}).}
\label{3crvvm1}
\end{figure}

To~model~the~luminosity dependence of the parent-source evolution
we adopt exponential luminosity-dependent-density evolution
({\it cf} `model 4b' of \pcite{wal80a}) such that the
evolution function $F(P,z)$ has the form 

\begin{equation}
F(P,z) = \exp M(P) \tau(z)
\label{eq:fpz}
\end{equation}

\noindent where $\tau(z)$ is the look-back time in units of the Hubble
time. For Einstein-de-Sitter ($\Omega$=1) geometry this is given by 

\begin{equation}
\tau(z) = (1 - (1 + z)^{-1.5}). 
\label{eq:tau}
\end{equation}
\noindent
Additionally we apply a redshift cutoff to the populations
mirroring the observed behaviour of the powerful radio sources
at high redshift \cite{sha96}.  The evolution
function is modified such that
the evolution peaks at $z_c/2$ and declines to zero at the cut-off
 redshift $z_c$: \\

$F = F(P,z)$ for $z \le z_c/2$, \\

$F = F(P,z_c - z)$ for $z_c/2 < z \le z_c$ and \\

$F = 0$ for $z > z_c$. \\

\noindent The evolution rate $M$ is set between 0 and $M_{max}$ dependent
on radio power $P$: \\

$M(P) = M_{max}  \frac{\log_{10}P - \log_{10}P_{1}}{\log_{10}P_{2} - 
\log_{10}P_{1}}$
\hspace*{0.3in} for $P_{1} \leq P \leq P_{2}$, \\

$M(P)$ = 0  for $P < P_{1}$, {\it i.e.} no evolution of radio 
sources of radio power less than $P_{1}$, \\

and $M(P) = M_{max}$  for $P > P_{2}$, {\it i.e.} sources
of radio power greater than $P_{2}$ undergo maximal evolution. \\

\subsection{Determining the evolution}

\subsubsection{Deriving the LRLF} 

\label{dlrlf}

Luminosity distributions were compiled for the
powerful radio sources from the
complete 3CRR sample, comprising
173 sources with $S_\fb \geq$ 10.9 Jy.
In this analysis the 10 flat-spectrum sources 
($\alpha_\fb^\fg 
> -$0.5, where $S \propto \nu^{\alpha}$) are excluded so the
remainder are steep-spectrum sources,
`uncontaminated' by the effects of Doppler beaming.

\begin{figure}
\vspace{3.0in}
\includegraphics{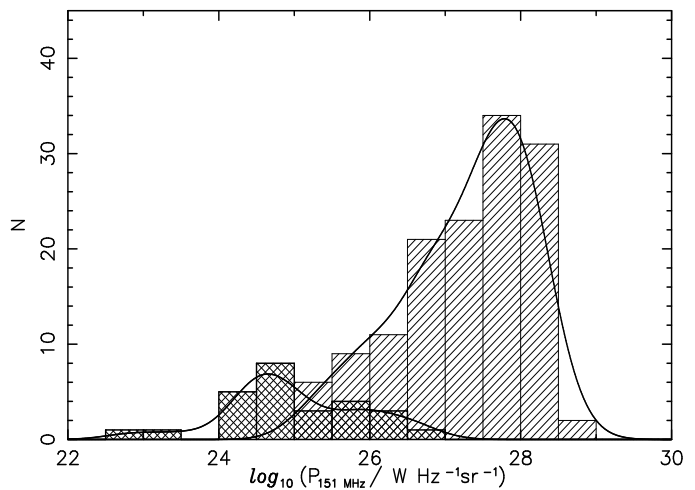}
% from cam_home\ldss3cr(4)
\caption{Luminosity distributions for the 26 FRI (cross-hatched)
and 137 FRII (hatched) radio sources in 3CRR.
The smoothed master luminosity distributions are over-plotted and
were derived with $\sigma$=0.35 in $\log_{10} (P_\fa)$ as discussed
in the text.}
\label{lds3c}
\end{figure}

The 178-MHz flux densities were translated to 151~MHz using
a single spectral index $\alpha^\fb_\fa=-$0.75,
the mean value for the 3CRR steep-spectrum sample. 
Separate luminosity distributions were compiled for the FRI and FRII
steep-spectrum parent populations using the morphological classifications
compiled by Laing et al. (1983) and by R Laing (private
communication). The total sample comprises 
26 FRI and 137 FRII sources.
The unbinned data were smoothed to give a master
luminosity distribution for each population by convolving each of the
unbinned luminosities
with a Gaussian curve of unit area, varying $\sigma$ until the sum
of all contributions is smooth whilst preserving the `real' 
features of the distribution.  The binned and master luminosity 
distributions for the two populations are shown in Figure~\ref{lds3c}. 

The local radio luminosity functions for
the FRI and FRII populations are then
determined from equation~\ref{eq:rhono}, using the
evolution function described by equation~\ref{eq:fpz}.
However, the scarcity of
radio sources in the 3CRR sample with radio powers below
$\log_{10} P_\fa$ = 10$^{24}$ W Hz$^{-1}$ sr$^{-1}$ yields an incomplete
LRLF at these radio powers. This incompleteness is resolved by
incorporation of the LRLFs determined for {\it (i)} the `local' E/S0
galaxies \cite{con84} considering them as part of the
FRI population,  and 
{\it (ii)} the starburst galaxies, with the LRLF as derived by
\scite{row93} and the evolution function
of \scite{sau90}. These starburst sources were folded into
the space density analysis with the set evolution function \\

$F_{star}(z) = \exp Q \tau(z)$ \\

\noindent with $\tau(z)$, the look-back time, as given in
equation~\ref{eq:tau}. The evolution function $F_{star}$ describes pure 
luminosity evolution as follows: \\

$Q$ = 3.1 for $0 \leq z \leq 2$, \\

$F_{star}(z) = F_{star}(z=2)$ for $2 < z \leq 5$, \\ 

and $F_{star}(z) = 0$ for $z > 5$. \\

\begin{figure}
\vspace{3.0in}
\includegraphics{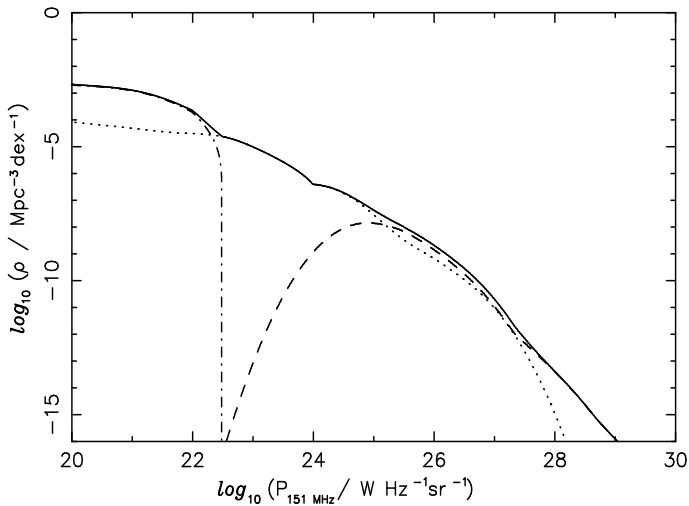}
% from cam_home\ rhop2rlf(3) 
\caption{Derived LRLF at 151 MHz. 
The solid line is the total LRLF with 
constituent populations FRI (dotted), FRII (dashed) and
starburst galaxies (dot-dashed). }
\label{lrlf1}
\end{figure}

\noindent
The total local radio luminosity function is
then the sum of the LRLFs for each of the three populations (FRI, FRII
and starburst galaxies).  The
derived LRLF (Figure~\ref{lrlf1}) shows that the 
transition between the FRI and FRII populations is gradual
and is not a simple transition at some radio power.  This is
also illustrated in Figure 4 where there are a number of 
FRIs in the 3CRR sample with $P_\fa > 10^{25}$ W Hz$^{-1}$ sr$^{-1}$ .
The luminosity functions were tapered
to avoid discontinuities, with the result that the space densities
for low-power FRIIs ($\log_{10} P_\fa <$ 23.5) and high-power
FRIs ($\log_{10} P_\fa >$ 27.5) are insignificant. 
We find good agreement with other 
LRLFs determined independently.   At the low-power end
the total LRLF agrees well with those determined by
\scite{dun90,sad89,aur77}. 
This agreement suggests that our FRI LRLF is reasonably well
defined. This definition is important as low-power FRI
sources are far more numerous locally than
the higher-radio-power FRI found in the 3CRR sample, and at low flux
densities their contribution is substantial.

The model FRI and FRII LRLFs from the Milne ($\Omega$=0) version of our 
successful model fit are in agreement
with the results of \scite{urr95}, expected as 
both analyses find little or no evolution of the FRI population.
However, there is a major discrepancy between the two FRII LRLFs.
Urry and Padovani adopted pure luminosity
evolution for the FRII population, which predicts a high local space
density of low-power FRIIs.  According to Urry and
Padovani's LRLF there should be similar local space densities 
of FRIs and FRIIs. Much smaller local space densities for FRIIs are
observed; \scite{led96} find only 6\% of 
FRIIs in a local survey of Abell clusters ($z <$ 0.09).  
However, over-estimation 
of the LRLF for low-power FRIIs does not affect the  
analysis performed by Urry and Padovani
which concentrates on fitting the RLF for the high-power
quasar population. 

\subsubsection{Fitting the 151-MHz source count}

\label{fitev}

The parameters in the evolution functions for FRI and FRII sources were
optimized by
performing a $\chi^2$-minimization between the observed source count
at 151 MHz and the total count predicted by the
differentially-evolving RLF.  The minimization was achieved using the 
{\it AMOEBA} downhill simplex method in multi-dimensions
\cite{pre92}.
%The contributions from each source
%type are handled individually (section~\ref{emethod}) with 
The unique set of evolution parameters thus derived for each parent
population is shown in Table~\ref{params}. 

\begin{figure*}
\vspace{4.0in}
\includegraphics{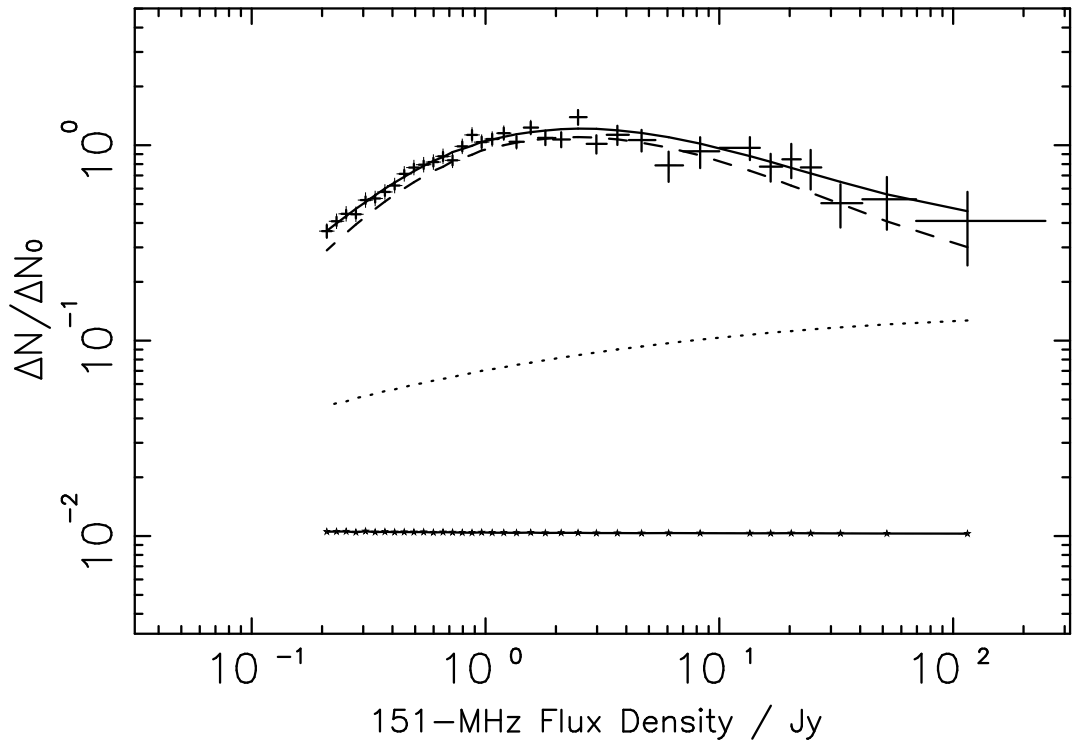}
% from cam_home\chi1
\caption{Source count fit at 151 MHz.
Data points represent the observed differential source count at 151 MHz
constructed as described in the text.
The model produces contributions from the three
populations, the FRII sources (dashed), the FRI sources
(dotted) and the starburst galaxies ($\star$).
The total model count is shown as a
solid line. All counts are shown in relative differential form with $N_{0} =
2400 (S_\fa)^{-1.5}$ sr$^{-1}$.}
\label{fite}
\end{figure*}

The 151-MHz source count used in this comparison (Figure~\ref{fite})
consists of 
the count from the 6C survey \cite{hal88} and the
count from the 3CR survey from
the 3CRR catalogue at 178 MHz \cite{lai83},
transposed to 151~MHz using a single spectral index of $-$0.75.

\begin{table}
\caption{Evolution functions used in the source count fit.}
\label{params}
\begin{tabular}{cccc}
           &  Evolution    &   Fixed & Evolution \\
Population & function & evolution ? & parameters \\
\hline
\\
 FRI  & exp LDDE   & No &  $M_{max}$, $z_c$, $P_1$, $P_2$ \\ 
\\
 FRII & exp LDDE  & No  & $M_{max}$, $z_c$, $P_1$, $P_2$ \\
\\
starburst & exp PLE & Yes & $Q$=3.1 \\
\hline
\end{tabular} \\
\end{table}

\begin{figure}
\vspace{3.0in}
\includegraphics{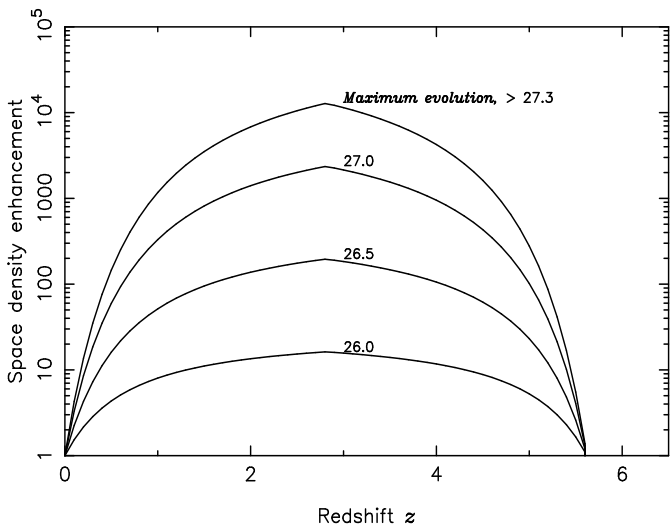}
% from cam_home\evolhis2
\caption{Space density enhancements as determined from optimized model
parameters for a range of $\log_{10}(P_\fa)$ values as shown.}
\label{ophse}
\end{figure}

The set
of parameters given in Table~\ref{fitet} provides a good fit to the
observed source count. This fit
has {\it strong} cosmic evolution of the FRII population coupled with
{\it no} evolution of the FRI population. The fit to the differential
count is shown in Figure~\ref{fite}.
For the most powerful FRII sources 
this model predicts a space density enhancement at $z$=2.8 of
$> 10^4$ that of the local space density
(Figure~\ref{ophse}).
That the FRI population
undergoes little or no evolution
is a {\it requirement} of the 151-MHz
source count fit: any significant evolution
produces an excess of faint sources, in disagreement with the decline
of the differential source count towards lower flux densities.
The successful fit also indicates a redshift
cut-off in the FRII population, with
the fit with $z_c$=5.62 superior to that with $z_c=\infty$ at the
99.9\% level of significance.  As a result the model
reproduces the `quasar epoch',
reflected as a peak in the FRII space density (Figure~\ref{ophse})
around $z$=2--3, as seen 
for powerful flat-spectrum quasars \cite{sha96}.

\begin{table}
\caption{Fitted evolution parameter values.}
\label{fitet}
\begin{tabular}{cccc}
           &     & \multicolumn{2}{c}{Chi-square test} \\
Popu- & Evolution parameter values & $\chi^{2}_{min}$ & $\nu^{\dagger}$ \\
lation \\
\hline
\\        
 FRI &  $M_{max}$=0.0, $z_c$=5.0 \\
     &  ($P_1$ \& $P_2$ not used given $M_{max}$=0.0) \\
\\
 FRII & $M_{max}$=10.93, $z_c$=5.62, \\
      & $P_1$=25.44, $P_2$=27.34 \\
\\
\hline

&   {\it best fit}        & 30.73 & 33 \\
\hline
\end{tabular}
\vspace{0.1in} 
$^{\dagger}$ degrees of freedom.
\end{table}

\section{Radio-source beaming in the unified scheme}

According to the dual-population unified scheme, 
flat-spectrum quasars and BL Lac-type sources are the
steep-spectrum FRII and FRI
sources aligned so that radio
axes are close to our line-of-sight. The result is that
the flat-spectrum emission from the approaching radio jet\footnote{It is
the {\it base}
of the approaching jet which is flat-spectrum. Alternative scenarios
are that the relativistic emission is from
(i) the stationary core of the radio source which is optically thick
at the observing frequency or (ii) synchrotron plasma emanating
from the central object.  All three scenarios in agreement with the
results here.} is Doppler-boosted or `beamed'.  This Doppler boosting can
result in the jet emission dominating the extended lobe emission. 

Having derived the space density and its epoch-dependence for the parent
populations, the low-frequency counts and identification statistics
are satisfied. Translating the count of these
steep-spectrum objects to 5~GHz reveals the shortfall at high flux
densities in particular which must be there: no flat-spectrum
sources have yet been included and such sources constitute more than
half of those found in cm-wavelength surveys. To provide this
contribution we `beam' the parent populations, orienting objects randomly
to mimic the effects of the Doppler-enhanced radiation when lines-of-sight
and ejection axes come into close coincidence. The parameters required to
describe the beamed
emission for any source are the spectral index of the core/jet emission,
the Lorentz factor $\gamma = (\sqrt(1 + (v/c)^2)^{-1}$ for the bulk motion
in the jets, and the 
the intrinsic ratio $R_c$ of core-to-lobe radio luminosity. We
adopt a simple model of the Doppler beaming for each of the two
populations, with the beaming parameters 
varied to produce the best statistical fit to the 5-GHz source count.

\subsection{The Doppler beaming parameters}

\begin{figure}
\vspace{3.0in}
\includegraphics{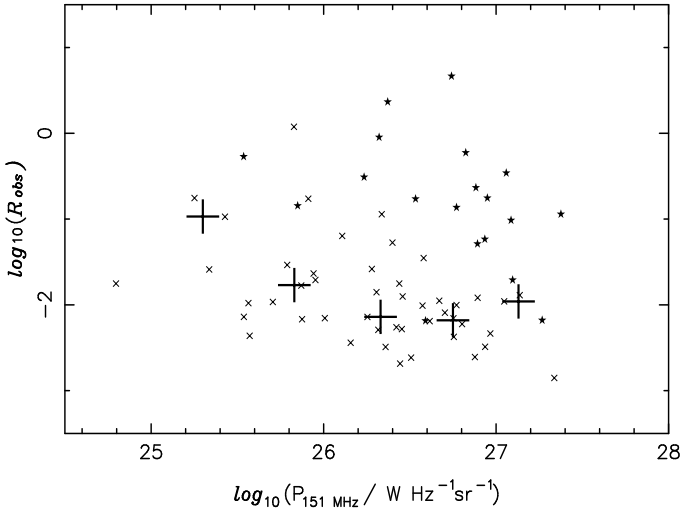}
% from cam_home/cam_save_prgs/core3cr
\caption{Observed core-to-extended flux
ratios, $R_{obs}$, for high-excitation FRII
sources in 3CRR, 47 galaxies ($\times$)
and 19 quasars ($\star$). Median $R_{obs}$ values for the galaxies,
{\it i.e. those sources which are unbiased by Doppler beaming} are shown
as ($+$). From data compiled by R Laing.}
\label{friicore}
\end{figure}

We chose to model the Doppler beaming of the FRI and FRII populations
with a range of intrinsic core-to-extended flux ratios
(distributed normally about a median value) together with a single
Lorentz factor for each parent population. There is an additional
complication for the FRI population, for which the intrinsic ratio for the
is a function
of radio power with significant scatter about a median value \cite{der90}.
In contrast, there is little evidence that the ratio
correlates with radio power for the FRII population.
Figure~\ref{friicore} shows the observed ratio for 47 high-excitation FRII
galaxies from the 3CRR sample.  The median ratio appears uncorrelated to
total radio power, with a significant scatter around the median values. 
The same result has been seen \cite{mor97} in narrow-line radio
galaxies from the 2-Jy sample \cite{wal85}. 

For the FRI population we adopt the following function
from the observed ratio for B2 galaxies \cite{der90}: \\

$\log_{10} R_{med} = -0.55 \log_{10} (P_\fa) + 10.78$. \\

\noindent Where $R_{med}$ is the median ratio for FRIs of radio power $P_\fa$.
The intrinsic core-to-extended flux ratio $R_c$ is 
then determined as a normal distribution about $R_{med}$ with 
 $\sigma$ = 0.45$R_{med}$. 
For the FRII population we reflect the observed scatter in 
the intrinsic ratio $R_c$ by simply distributing it normally about a
single median value $R_{med}$, with $\sigma$ = 0.45$R_{med}$. 
The derived intrinsic ratio, $R_c$, 
is frequency-independent,  
apportioning the total source flux density between
the extended and core emission.

\subsection{Modelling the beamed products}

\label{modelb}

\begin{figure*}
\vspace{3.5in}
\includegraphics{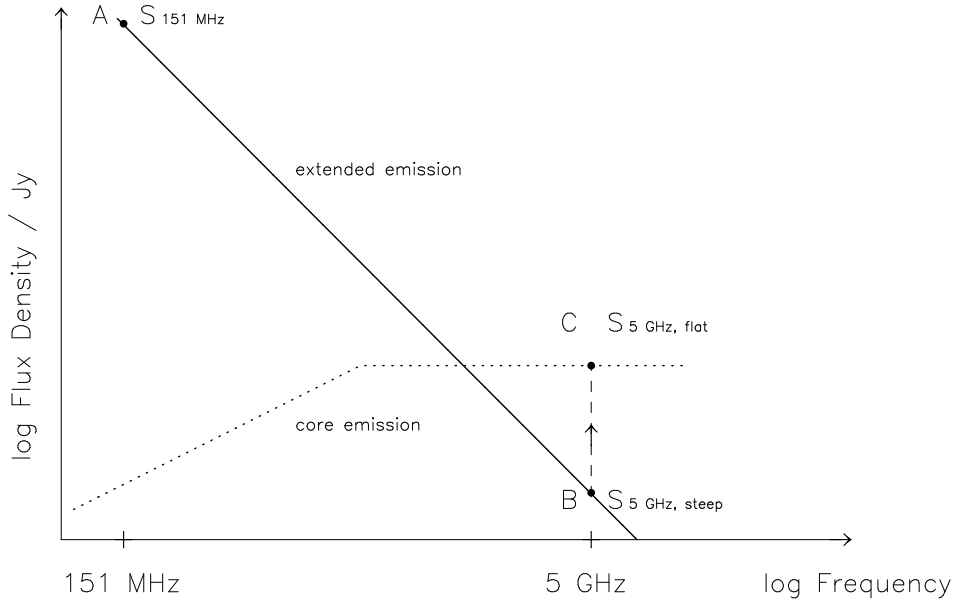}
% from pg_exam/seminars/fig7b ***CODE LOST ?***
\caption{Modelling the flat- and steep-spectrum components
of the FRI and FRII sources at 5~GHz. The contribution from the steep-spectrum
component at 151 MHz (A) is transposed to 5 GHz (B) using
$\alpha^\fe_\fa = -$0.75 (solid line).  
If the radio axis lies close
to the line-of-sight ($\theta \le \theta_{c}$)
the core emission (dotted) dominates
and the source appears `flat-spectrum'; the
total flux density of the source
is $S_{5 {\rm \thinspace GHz}, steep} + S_{{\rm \thinspace 5GHz}, flat}$.} 
\label{mfs}
\end{figure*}

The FRI and FRII populations give rise to
beamed and unbeamed sources with the {\it observed} 
core-to-extended flux ratio, $R_{obs}$, determining
whether a particular source is observed as 
flat- or steep-spectrum. The procedure is illustrated in
Figure~\ref{mfs} and described in detail below. Throughout
all flux densities are K-corrected to observed-frame
values using $\alpha=-0.75$.

\begin{enumerate}

\item A single spectral index, $\alpha^\fe_\fa  = -$0.75
is applied to the 151-MHz flux density to determine the flux density
of the steep-spectrum contribution at 5~GHz.
This represents the emission from the
extended lobes of the source.  The value of $-$0.75 is adopted as it is
the mean spectral index of the steep-spectrum 3CRR sources
between 178~MHz and 750~MHz \cite{lai80}.  We extrapolate
this index to 5~GHz as there is no
evidence that the spectrum of the steep-spectrum
components in 3CRR steepens significantly between 178 MHz and 2.7 GHz
\cite{lai80}. 

\noindent This steep-spectrum component flux density,
$S_{5 {\rm \thinspace GHz}, steep}$, is \\

$S_{5 {\rm \thinspace GHz}, steep}$ = $S_\fa . (5000 /151)^{-0.75}$, \\

\noindent shown as line A to B in Figure~\ref{mfs}. \\ 

\item The emission from the flat-spectrum core of the source
is calculated assuming that the source
comprises a pair of oppositely-directed relativistic jets
of bulk plasma velocity $\beta c$.
The ejection axis of these jets is
aligned at some random angle $\theta$
($0^{\circ} \leq \theta \leq 90^{\circ}$)
to our line-of-sight.  The
observed core-to-extended flux
ratio, $R_{obs}$, determines the degree of beaming observed
as follows: \\

$R_{obs} = R_{c} \Delta$   \\

\noindent where $\Delta$ is the sum of the Doppler enhancement from the
forward- and counter-jets of the source: \\

$\Delta = \delta_{f}^{p} + \delta_{c}^{p}$. \\

\noindent with the Doppler factor for the forward-jet, $\delta_f$, as \\

$\delta_f = (\gamma(1 - \beta \cos \theta))^{-1} $ \\

\noindent and for the counter-jet, $\delta_c$, as \\

$\delta_c = (\gamma(1 + \beta \cos \theta))^{-1}$ \\

\noindent where $\beta = v/c = (1 - \gamma^{-2})^{1/2}$. \\

\noindent For radio emission comprising continuously-ejected
plasma, $p = 2 - \alpha_{flat}$.  We adopt a value for the spectral
index of the core emission, 
$\alpha_{flat}$, of 0.0 as this is the mean spectral index
$\alpha_{408 MHz}^{1.4 GHz}$ of the core components of the
B2 sources \cite{der90}. \\

\noindent
Therefore the observed core-to-extended flux ratio is given by  \\

$R_{obs} = R_c ((\gamma(1 - \beta \cos \theta))^{-2} +(\gamma(1 + \beta \cos
\theta))^{-2}).$ \\

\noindent
The flux density of the beamed contribution,
$S_{5 {\rm \thinspace GHz},flat}$, is represented by line B to C in
Figure~\ref{mfs} and is simply related to the 
steep-spectrum flux density by
$S_{5 {\rm \thinspace GHz}} = R_{obs} . S_{5 {\rm \thinspace GHz}, steep}$.  \\

\item A source is counted as `flat-spectrum' for
values of  $R_{obs}$  large enough such
that $\alpha^\fe_\fd \ge -$0.5.
This determines the value of $R_{min}$, the lowest $R_{obs}$ value
at which a source is counted as flat-spectrum, as follows: \\

\noindent
The spectral index between 2.7 GHz and 5 GHz is given by

\[ \alpha^\fe_\fd = \frac{\log_{10}\frac{S_{5 {\rm \thinspace GHz}, total}}
{S_{2.7 {\rm \thinspace GHz}, total}}} 
{\log_{10}\frac{\nu_\fe}{\nu_\fd}}.  \]

\noindent
For $\alpha_{flat}$=0.0 the flat-spectrum components
have the same flux density at 2.7 GHz and 5 GHz, {\it i.e.}
$S_{2.7 {\rm \thinspace GHz}, flat}$ = $S_{5 {\rm \thinspace GHz}, flat}$ 
and for $\alpha_{steep}$=$-$0.75 the steep-spectrum components
at 2.7 GHz and 5 GHz are related as
$S_{2.7 {\rm \thinspace GHz}, steep}$ = 1.6 $S_{5 {\rm \thinspace
GHz},steep}$. Thus we determine $R_{min}$ as having the value of 0.66. 

\end{enumerate}

The largest angle between the jet and the line of sight for the source
to appear flat-spectrum occurs at the the critical angle $\theta_{c}$
when $R_{obs}$ = $R_{min}$.  Thus
$\theta_{c}$ = $\delta_{min} =  \delta(\theta_{c})$  so that $\theta_{c}$
is determinable from
\begin{equation}
\cos (\theta_{c}) = \frac{1}{\beta} -
\frac{1}{\beta \gamma} (\frac{R_{c}}{R_{min}})^{p^{-1}}, \\
\end{equation}

\noindent
assuming that the beaming due to the counter-jet is negligible.  
In fact our successful model finds that $\theta_{c} <$ 10$^{\circ}$ 
and the assumption is justified. 

Thus the only unknown parameters in the 5-GHz
source count fit are values for $\gamma$ and $R_{med}$ for each
of the FRI and FRII populations.

\subsection{Fitting the 5-GHz source count}

\label{fit5b}

The  5-GHz source count is constructed from survey data
as detailed in Table~\ref{hifreq} and is
well defined across a wide flux-density range. 
At 5 GHz, the beamed products of the FRI and FRII sources make
a highly significant contribution to the source count; for flux
densities above 0.5 Jy, some 55\% of sources are `flat-spectrum'
\cite{pau78}.

\begin{table*}
\caption{5-GHz source count data.}
\label{hifreq}
\begin{tabular}{cccl}
Flux density & Survey & Survey \\
range $S_\fe$ & instrument(s) & name & Reference\\
\hline
\\
10 Jy - 100 Jy & NRAO + MPIfR & strong sources &  \scite{kuh81} \\
1.5 Jy - 10 Jy & Green Bank 91m & 87 GB & \scite{gre91} \\
0.5 Jy - 1.3 Jy & NRAO + MPIfR & S4 & \scite{pau78}\\
67 mJy - 0.5 Jy & NRAO 300ft & 6cm survey & \scite{dav71} \\
10 mJy - 55 mJy & MPIfR & deep selected regions & \scite{pau80}\\
1.5 mJy - 8.5 mJy & VLA & E/S0 galaxy & \scite{wro90} \\
1.36 mJy - 6.00 mJy & VLA & Lynx-2 area & \scite{don87} \\
16 $\mu$Jy - 60 $\mu$Jy & VLA & DEEPS2 & \scite{fom91} \\
\hline
\end{tabular}
\end{table*}

To define the populations in accordance with our earlier discussion of
the FRII beamed products, we split the FRII parent sources into
high- and low-excitation types using a
simple linear function of $\log_{10}(P_\fa)$ as suggested by
\scite{lai94} and \scite{bar94}. The fraction of
low-excitation FRIIs was set at 50\% at $\log_{10}P_\fa =$ 25.0,
declining to zero at $\log_{10}P_\fa = $ 27.0).

\begin{table*}
\caption{Radio source types at 5 GHz.}
\label{hipops}
\begin{tabular}{clcc}
   &                              &        & Radio \\
   &   Source population/evolution/ & Beamed & spectrum \\
Ref$^{\sharp}$ & beamed \& unbeamed products & at 5 GHz ? &
$\alpha^\fe_\fd$ \\
\hline
\\
{\it FRII sources} & tapered exp LDDE, $F(P,z) = \exp M(P) \tau(z)$ \\
& {\it evolution parameters as determined in section 3}  \\
\\
1 & High-excitation radio galaxies \& quasars (class A) & no & steep \\
2 & Quasars & yes & `flat' \\
3 & Low-excitation radio galaxies (class B) & no & steep \\           
4 & BL-Lac type objects & yes & `flat' \\
\\
\hline
\\           
{\it FRI sources} & {\it  non-evolving as determined in section 3} \\
\\
5 &  FRI radio galaxies (class B)& no & steep \\
6 &  BL-Lac type objects & yes & `flat' \\
\\
\hline
\\
{\it low-power}  & exp PLE, $F(z) = \exp Q \tau(z)$  \\
& {\it PLE as described in section 3} \\
\\
7 &  starburst galaxies & no & steep \\
\hline
\\
\end{tabular}
\vspace*{0.05in}
\\ {\it $\sharp$ These numbers are used to reference the source types 
in Figures~\ref{fitb} to~\ref{pop14}.}
\end{table*}

The source count fit was carried out as for the 151-MHz count, this time
incorporating the beamed products of the FRI and FRII sources
by randomly aligning the sources with respect to our line-of-sight.           
The source alignment angle, $\theta$, for each of
the FRI and FRII contributions was  generated using
NAG pseudo-random number routines.
The best-fit beaming parameters were determined again using the
{\it AMOEBA} downhill simplex method,
minimizing $\chi^{2}$ evaluated between the observed and model source
counts at 5~GHz.  

A good fit to the 5-GHz source count was found for
the parameter values given in Table~\ref{paramsb}.

\begin{figure*}
\vspace{4.0in}
\includegraphics{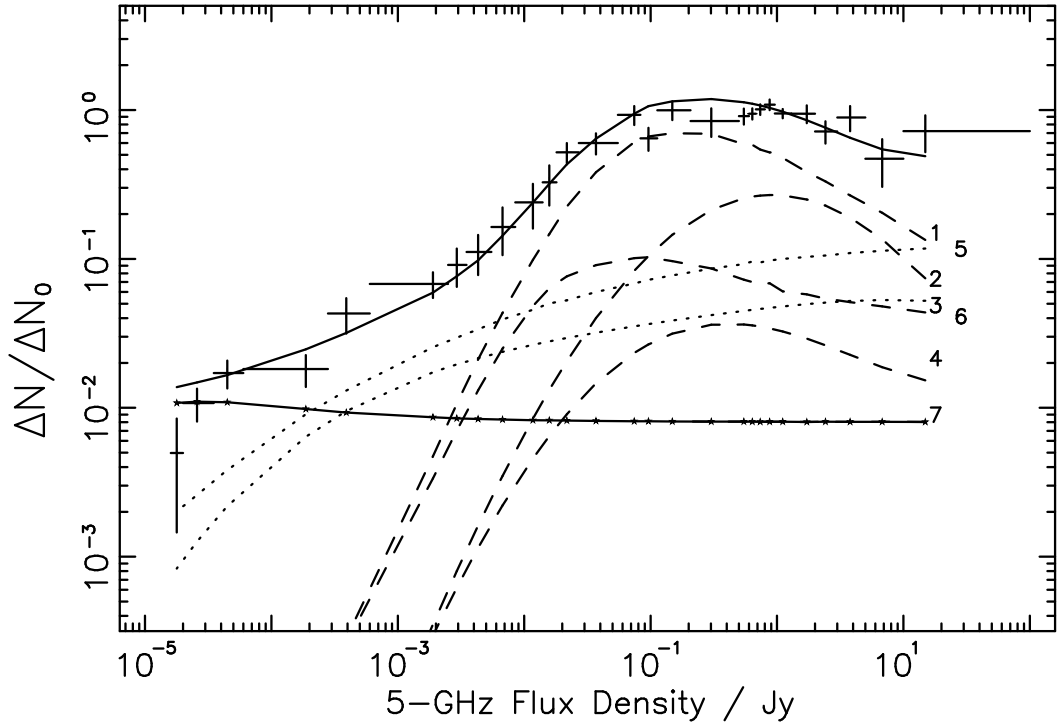}
% from cambridge evol/chi6 (usyd no NAG routine)
\caption{Source count fit at 5 GHz. Data points represent the 
differential source count at 5~GHz from data given in 
Table~\ref{hifreq}.
The model produces contributions from seven source types as
described in Table~\ref{hipops}.
All counts are shown in relative differential form with $N_{0} = 60
(S_\fe)^{-1.5}$ sr$^{-1}$. }
\label{fitb}
\end{figure*}

The observed source count and the count calculated for the
optimum fit are shown in Figure~\ref{fitb}.  Here the
sub-populations contributing to the count are split
into their beamed and unbeamed products as detailed
in Table~\ref{hipops}.  The contribution from
the beamed sources peaks at high flux densities
(0.1 $< S_\fe <$ 10 Jy), broadening the `evolution bulge' of the
count. Thus the unification scheme provides a
straightforward explanation of {\it why} the source-count `plateau'
widens with survey frequency. It is the
Doppler-beaming of  the FRII population which has
the primary impact in this respect. The combination
of zero evolution with milder
Doppler beaming for the FRI population results in a monotonically-decreasing 
count of beamed and unbeamed FRIs throughout the flux density
range as shown.

\begin{table}
\caption{Fitted beaming parameter values.}
\label{paramsb}
\begin{tabular}{lccc}
            &   & \multicolumn{2}{c}{Chi-square test} \\
Population  & parameter values & $\chi^{2}$ &
$\nu^{\dagger}$ \\
\hline
\\
FRII & $\gamma$=8.5, $R_{med}$=0.01 \\
      &  $\theta_{c}(R_{med})$=7\fdg1 \\
\\
FRI & $\gamma$=15.0 \\
      & $R_{med} \propto P_\fa^{-0.55}$ \\
\\
\hline
& {\it best fit} & 32.98 & 25 \\
\hline
\end{tabular}
\vspace{0.05in}  
\\ $^{\dagger}$ degrees of freedom.
\end{table}

\section{Testing the model}

The substantial predictive power of the model provides
direct tests of the dual-population unified scheme.

\subsection{The beaming models}

\subsubsection{Jet speeds; values of the Lorentz factor $\gamma$}

The single most successful aspect of the model is that
values of $\gamma$ determined from the optimization process 
agree with those determined from estimates made from 
(i) VLBI observations of
superluminal sources and (ii) synchrotron self-Compton models for the
observed and predicted X-ray flux.  

VLBI studies of superluminal sources measure the apparent velocities
of recognisable features in radio sources.  Results from a large
($>$ 100) sample of superluminal sources show that, assuming that
the bulk and pattern speeds are the same,
$\gamma_{min} \sim$  4 are commonly found although there
is a wide spread of values from $2 \leq \gamma \leq 20$
\cite{ver95}.

The synchrotron self-Compton model estimates the
Doppler beaming from the observed and predicted X-ray flux
densities of radio sources.  For a sample of $\sim$100 sources
a {\it lower} limit for the bulk Lorentz factor of $\sim$10
is found with a viewing angle $\sim$8$^{\circ}$ for all
superluminal sources.  This analysis also indicates that
the core-to-extended flux ratio has a significant scatter
\cite{ghi93}.

In addition to the above, $\gamma \sim 10$ successfully accounts for the
non-detection of a counter-jet in highly-aligned objects
such as 3C\,273 \cite{dav91}.  

\subsubsection{The observed core-flux : extended-flux ratio}

The observed core-to-extended flux ratios, $R_{obs}$, 
span a wide range of values. 
The most lobe-dominated sources, those unaffected by
Doppler beaming, have ratios as low as
$R_{obs}\sim 10^{-5}$  ({\it e.g.} OD-159 from the 2-Jy sample,
\pcite{mor93}). In contrast
the most core-dominated sources are heavily
Doppler-beamed and can have
$R_{obs}\sim10^{3}$ ({\it e.g.} PKS B0400+258,
\pcite{mur93}). 

Our assumption of a normal distribution for $R_c$ about $R_{med}$ 
results in a wide range of  $R_{obs}$  for both
FRI and FRII beamed products (Figures~\ref{frirobs}
and~\ref{friirobs}), and provides a smooth transition 
between the parent objects and their beamed sources.
The $R_{obs}$ values for BL Lac-type sources are predicted to
extend to higher values due to the large range of intrinsic
$R_c$ values. The limited data in complete samples appropriate
for comparison are in approximate agreement.

\begin{figure}
\vspace{3.0in}
\includegraphics{robsi5C.figps}
% from evol/rhop2fsl(2)
\caption{Model $R_{obs}$ distribution for the
flat-spectrum BL-Lac type sources (cross-hatched), plus
steep-spectrum FRI and low-excitation FRII sources
(hatched), over 1 sterad  with $S_\fe \ge$~0.1~Jy. }
\label{frirobs}
\end{figure}

\begin{figure}
\vspace{3.0in}
\includegraphics{robsii5C.figps}
% from evol/rhop2fsl(2)
\caption{Model $R_{obs}$ distribution for (i) flat-spectrum
quasars (hatched), (ii) steep-spectrum quasars and broad-line radio
galaxies, {\it i.e.} high-excitation FRIIs with radio axes within
50$^{\circ}$ 
of the line-of-sight (hatched), and (iii) narrow-line radio galaxies, 
high-excitation FRIIs with
radio axes lying $ >50^{\circ}$ from the line-of-sight (solid fill),
over 1 sterad  and with $S_\fe \ge$ 0.1 Jy. }
\label{friirobs}
\end{figure}

\subsection{Multi-frequency source counts}

Figure~\ref{wbsc} shows the
observed differential source counts at a wide range of
radio frequencies together with the predictions of counts from the model.
The model successfully 
reproduces these source counts, qualitatively at least.
There are quatitative differences, most notably the
excess in the prediction at low
flux densities for 1.4 GHz. The most proabale explanation is that
our simple model of the
local luminosity function overestimates the space density for powers
below $\log_{10}(P_\fa)\sim$ 24.0.

\begin{figure*}
\vspace{6.0in}
\includegraphics{wbsc.figps}
% from src_cnts/cjcnts3 ** NOTE THAT FIRST SURVEY NOW USED AT 1.4 GHz **
\caption{Observed source counts at six frequencies with
data points from surveys as given in Wall (1994) and the
addition of the  VLA FIRST survey source-count points 
1 mJy $\le S_\fc \le$ 1 Jy \protect\cite{bec95}.
The predicted model counts are shown dashed.  All counts are
in relative differential form
$\Delta N / \Delta N_{0}$, where $\Delta N$ is the number of sources per
sterad with flux density $S_{\nu}$ between $S_{2}$ and $S_{1}$ and
$N_{0} = K_{\nu} S_{\nu}^{-1.5}$, the number of sources expected in a
uniformly-filled Euclidean universe. The values of 
$K_{\nu}$ are 2400, 2730, 3618, 4247, 5677 and 3738 respectively for the
six frequencies shown. The horizontal range bars show
the flux-density bin width $S_{2}$ to $S_{1}$, and the error bars
represent $\sqrt N$ uncertainties.  Polygons show the count estimates
from $P(D)$ (background-deflection) analyses.}
\label{wbsc}
\end{figure*}

\subsection{Core-dominated and broad-line fractions}

\begin{figure}
\vspace{3.0in}
\includegraphics{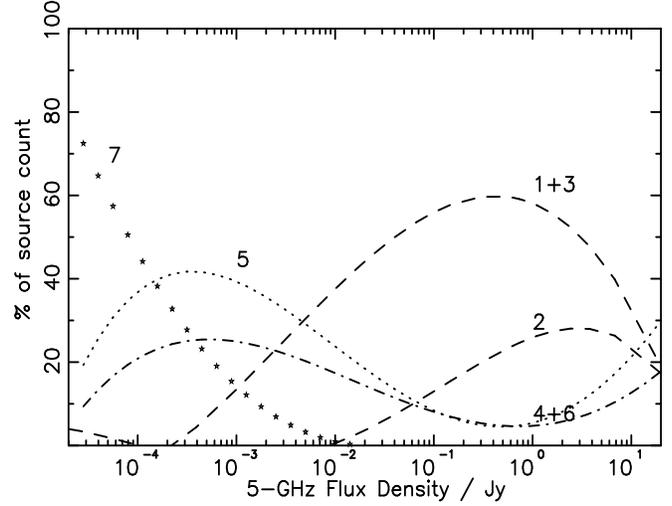}
% from populations/newpop5d
\caption{The predicted population mix at 5 GHz, with
populations as given in Table~\ref{hipops}.}
\label{pop5}
\end{figure}

\begin{figure}
\vspace{3.0in}
\includegraphics{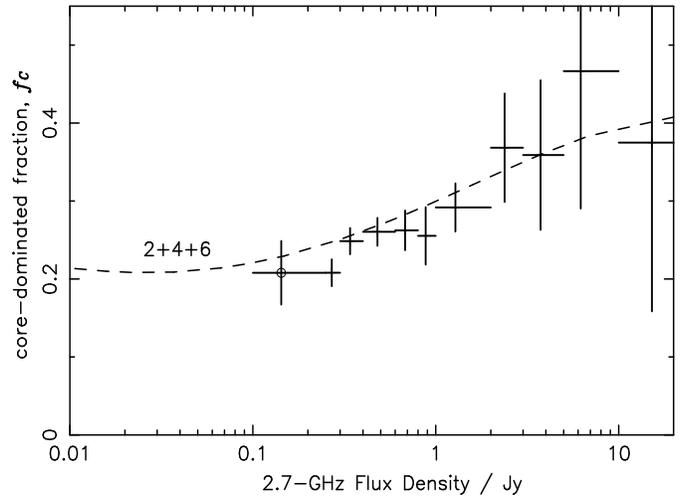}
% from populations/newpop2d
\caption{Core-dominated fractions at 2.7 GHz.
The dashed line is the model prediction of core-dominated
sources (BL\,Lacs plus quasars).
The data points are derived from two samples discussed in detail
by Wall \& Jackson (1997), section 3.1:
PKSCAT90 (+) with $S_{2.7 GHz}
\geq $ 0.25 Jy and  PSR ($\circ$) with 0.10 $\leq S_{2.7 GHz} < $ 0.25 Jy.
The error in $f_C$ is $\sqrt(N_{q})$ / (bin total).}
\label{qfrac}
\end{figure}

Figure~\ref{pop5} shows the population mix as predicted by the
model as a function of 5-GHz flux density.
The striking feature is the change in dominant population
moving from high flux densities (high-excitation FRII radio
galaxies and quasars) to low flux densities (FRI radio galaxies,
BL Lac-type sources and - eventually - starburst galaxies).
The domination of the high-excitation
FRII sources at high flux densities
produces a pronounced peak in the quasar population
between 0.5 and 10 Jy; the decline in quasar fraction
towards lower flux densities is in close agreement with observation
(Figure~\ref{qfrac}), as discussed by \scite{wal97}.

The change in fraction of broad-line sources as a function of flux 
density agrees with the best available determinations at 408 MHz (Figure 5
of \scite{wal97}.
Such a change is evidence for the unified scheme rather than
evidence against it as was suggested by \scite{sin96}. 
Despite its relative simplicity, the dual-population scheme shows that
tests of unification dealing with broad-brush proportions of
broad-line objects or flat-spectrum objects are far too simplistic;
the mix of sub-populations selected in each sample must be considered.
There must be similar reservations over elementary tests based on size
discrimination ({\it e.g.} \pcite{kap96}).

\subsection{The redshift distribution of flat-spectrum quasars}

\begin{figure}
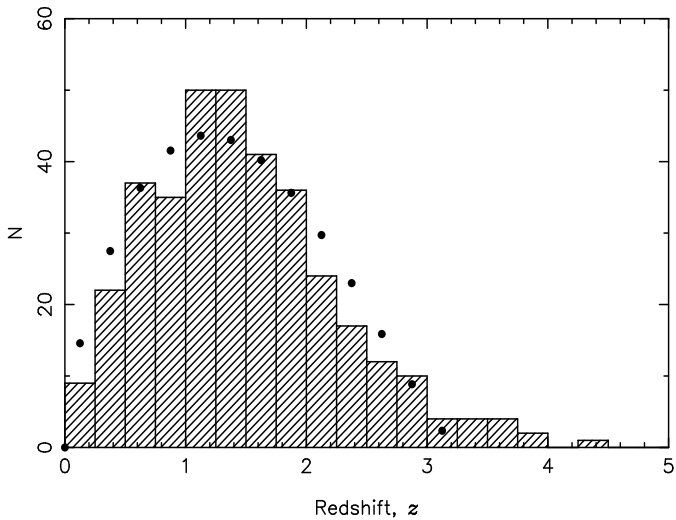

\vspace{3.0in}
\includegraphics{sampqzs.figps}
% from ids/pksplot3
\includegraphics{modelqzs.figps}
% from evol/rhop2b4q & plot27n3
\caption{The redshift distribution for 358 flat-spectrum quasars
from Shaver et al. (1996), shown with the model
prediction for 358 quasars selected with
$\alpha_\fd^\fe \ge$ $-$0.4 and $S_\fd \ge$~0.5~Jy.}
\label{sampqzs}
\end{figure}

A single sub-population redshift distribution, $N(z)$, provides a potent 
test of the model.  The flat-spectrum sample of \scite{sha96} 
comprises 444 sources, all of which are identified. Of these,
358 quasar redshifts were available to us.
The sample is moderately complete to
$S_\fd$= 0.25 Jy although a substantial proportion of the sample is
limited to $S_\fd$=0.6 Jy.
All sources were selected as flat-spectrum, with
$\alpha_\fd^\fe \ge$ -0.4 ($S \propto \nu^{\alpha}$).
The redshift distribution for the observed sample of
358 quasars is shown in
Figure~\ref{sampqzs},  along with the model prediction for
an equivalent {\it complete} flux-limited sample
($S_\fd \ge$ 0.5 Jy).
The model successfully replicates the observed $N(z)$
for the largest sample of spectroscopically-identified
quasars available. It should be emphasized that these data are
completely independent of the (low-frequency) data used to construct the
model.

\subsection{Radio surveys to mJy flux-density levels}

The new, large-area 
radio surveys ({\it i.e} FIRST, NVSS, WENSS \&
SUMSS) reach mJy flux-density levels and
will yield samples forming the basis of future
direct tests of the dual-population model.

\begin{figure}
\vspace{3.0in}
\includegraphics{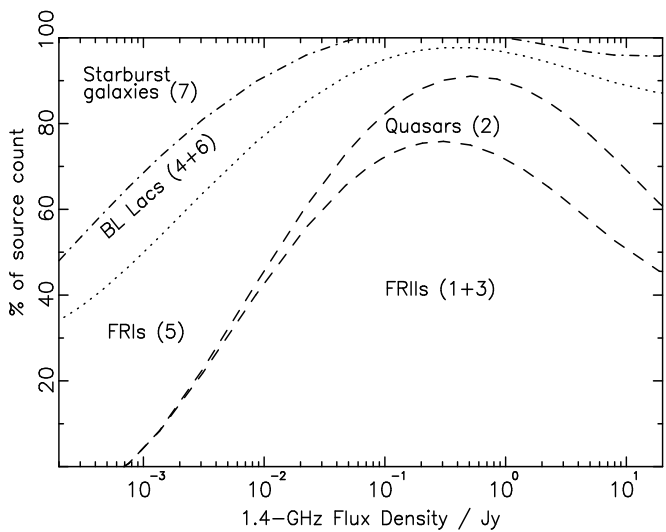}
% from populations/newpop5d
\caption{The predicted (integral) population mix at 1.4 GHz with the
populations as described in Table~\ref{hipops}.}
\label{pop14}
\end{figure}

Figure~\ref{pop14} shows the predicted integral population mix as a
function of 1.4-GHz flux density.
The dominant populations in the range 1 $\le S_\fc <$ 100 mJy
differ substantially from those at higher flux density. 
A direct consequence of the
dual-population model with its strong differential evolution is the
dramatic decline in prominence of the high-excitation FRIIs (and hence
broad-line objects) toward 1 mJy.

This change in population with flux density likewise
has major influence on the redshift
distribution of sources, as Table~\ref{tabzs} illustrates.
At 0.1 $\le S_\fc < $ 100 Jy the population
is predominantly FRII sources at high redshift, while
a very much more local distribution of sources appears at
1  $\le S_\fc < $ 100 mJy as the low-power FRIs and eventually the
starburst galaxies rise to prominence.

\begin{table*}
\caption{Population mix at 1.4 GHz.}
\label{popsat14}
\begin{tabular}{lccccccc}
\\
  & \multicolumn{2}{c}{FRII high-excitation} &
\multicolumn{2}{c}{FRII low-excitation} & \multicolumn{2}{c}{--- FRI ---}  &
Starburst \\
Flux density range  & BL \& NLRGs & Quasars & RGs &  BL Lac & 
RGs & BL Lac & galaxies \\
 &   (1)       & (2)     & (3)   & (4)    & (5) & (6) & (7) \\
\hline
\\
$S_\fc >$ 100 mJy  & 59\% & 15\% & 10\% & 4\% & 7\% & 4\% & 1\% \\
\\
1 $\le S_\fc \le$ 100 mJy & 7\% & 1\% & 4\% & - & 38\% & 30\% & 20\% \\
\\
\hline
\end{tabular}
\label{tabzs}
\end{table*}

\begin{figure}
\vspace{4.2in}
\includegraphics{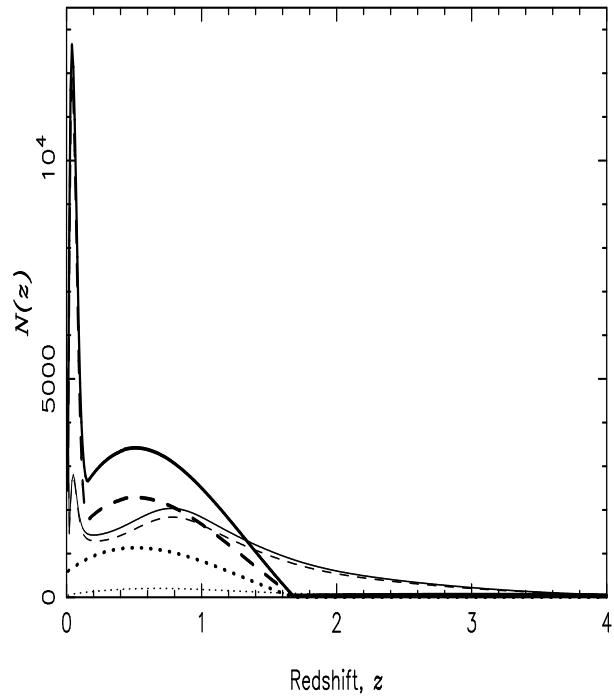}
\caption{The predicted $N(z)$ (per 0.01$z$ per sterad) for sources
with 1 $\le S_\fc <$ 100 mJy. Models are from
Dunlop \& Peacock (1990) (average of their models 1-7), 
shown as thin lines, and this paper, shown as thick lines.
Legend: \ldots flat-spectrum sources, - - - steep-spectrum sources
and solid line total of all sources. }
\label{comp14nz}
\end{figure}

Figure~\ref{comp14nz} shows a comparison between $N(z)$ distributions
for a flux density range  1  $\le S_\fc <$ 100 mJy.
The striking differences between our model prediction and the average N(z)
model
of \scite{dun90} are {\it (i)} the sharp spike at $z
\sim$ 0.1
due to the starburst galaxies in our model, a population not included
in the Dunlop \& Peacock analysis, {\it (ii)} the lower median redshift
distribution of sources in the prediction of the current model, due to the
dominance of the unevolving
FRI sources within this flux density range, and {\it (iii)} the
greater prominance of flat-spectrum sources in our model, the beamed
products of the low-power FRI sources. 
        
Although the model predicts unequivocal N(z) and population mix, {\it e.g.}
for flux densities in the range $1 < S_{1.4GHz} <100$ mJy, it has been
developed from consideration of {\it total} source counts at high and low
frequencies, but {\it population mix} only at low frequencies and high flux
densities, $S_\fb > 10.9$ Jy. Complete samples (with redshifts)
at the mJy level offer substantial scope for improvement or refinement of
the parameters.  For example, the sample at $S_\fd = 0.3$ mJy of
\scite{gru97} contains more broad-line objects than predicted. If
confirmed, this suggests that we have overestimated the the FRI LRLF at low
powers, a possibility already suggested in \S 5.2.  Tapering off the FRI
LRLF more rapidly at its higher radio powers would require a compensatory
broadening of the FRII LRLF, resulting in an increased proportion of FRII
objects at sub-mJy levels. 

\section{Conclusions}

We have described a paradigm in which evolution of extragalactic radio
sources is delineated for populations which are physically related through
orientation-dependent effects. The dual-population scheme is based on the
premise that the two parent populations are radio galaxies of FRI and FRII
morphologies, with distinctly different evolutions which we have determined
from analysis of counts and identification data at low frequencies. The
fundamental assumption is that all FRII and FRI radio galaxies have radio
lobes fed with relativistic beams. The beaming models by which these parent
objects appear as the quasars or BL\,Lac objects when lines-of-sight
coincide closely with radio axes were then developed by considering the
high-frequency counts and identifications, dominated by these beamed
products.
 
The model provides agreement with source-counts and with identification and
redshift data at high flux densities and at both high and low frequencies. 
It provides a natural explanation of the change in shape of source counts
with frequency. Furthermore, such a unified scheme provides a natural
explanation of why steep-spectrum sources show more pronounced cosmological
evolution than do beamed (flat-spectrum) sources of the same (apparent) 
power \cite{dun90}, a result anticipated by \scite{sch79}. 

One of the clearest successes of the model is that the beam speeds derived
from statistics of counts and identifications are in reasonable agreement
with those determined for individual sources from VLBI measurements.
Further it predicts a core-dominated fraction of sources as a function of
flux density in excellent agreement with observation.

Four further remarks are important with respect to the current model:

\begin{enumerate}

\item Although we chose LDDE to describe the evolution, it can be shown
that due to the shape of the RLF {\it some} models of LDDE are exactly
equivalent to PLE. Obviously a simple solution of PLE would be very
interesting with regard to physical models for {\it all} AGN evolution
(radio-loud and radio-quiet). 

\item The model fit of LDDE has the lower-radio power FRIIs undergoing
little or no evolution.  By virtue of the correlation between line-strength
and radio luminosity, these are inevitably of weak emission-line type. This
suggests that the weak-lined FRIIs underwent a similar evolution history as
FRIs; it is only the powerful strong-emission line FRIIs which evolved
strongly with cosmic epoch.  It might then be possible to define the two
populations in terms of optical emission-line strength rather than radio
morphology, the two emission-types perhaps representing different accretion
types. Such a view has important ramifications for the process of physical
evolution; and it may be that development of such a variant of the model
could result in a single source function describing the entirety of the
radio-source population. 

\item Although the simple beaming models adopted provide a good description
of the count data, there are other combinations of the Doppler beaming
parameters which work.  For example a very wide distribution of
Lorentz factors also fits the GHz-frequency source counts and this variant
was adopted by Urry and Padovani (1995).  Ultimately our selection of
parameters was made on the basis that the intrinsic core-to-extended flux
ratio is an observed quantity and that the model distribution of $R_{obs}$
values produced has the observed shape. In addition, the single $\gamma$
value fitted in our model can be interpreted as modelling the mean value
for the fastest material in the jets.  The critical angles inferred from
the successful models are $\sim$8$^{\circ}$ and $\sim$5$^{\circ}$ for the
FRI and FRII populations respectively.

\item It must be borne in mind that the current model has been developed 
fundamentally with only three data sets: a source count at 151 MHz, a
source count at 5 GHz, and a complete set of identifications and
redshifts at 178 MHz. It is clear that the model can be
improved. For example it predicts a low flux-density end of the 1.4-GHz
count which is too high, and it predicts perhaps too few broad-line objects
at low
flux densities. Both deficiencies could be rectified with  a local radio 
luminosity function for which the FRI component rolls off at somewhat
lower powers than the current one, compensated by extending the
FRII component to these lower powers.

\end{enumerate}

However in view of the limited data from which the scheme was developed,
its general success in describing much of the remaining data provides 
indication that the concept is substantially correct.

\bibliography{jw}

\bibliographystyle{mn}

\bsp

\label{lastpage}

\end{document}